\documentclass[prd,nofootinbib,showpacs,superscriptaddress,preprint]{revtex4}
\usepackage[T1]{fontenc}
\usepackage{amsmath,amssymb}
\usepackage{epsfig}
\usepackage{graphicx}
\usepackage[usenames,dvipsnames]{color}
\usepackage{slashed}
\usepackage[colorlinks,citecolor=blue]{hyperref}
\usepackage{pdfpages}
\usepackage{color}
\begin{document}
\title{Common Origin of Non-zero $\theta_{13}$ and Baryon Asymmetry of the Universe in a TeV scale Seesaw Model with $A_4$ Flavour Symmetry}

\author{Debasish Borah}
\email{dborah@iitg.ernet.in}
\affiliation{Department of Physics, Indian Institute of Technology Guwahati, Assam 781039, India}
\author{Mrinal Kumar Das}
\email{mkdas@tezu.ernet.in}
\affiliation{Department of Physics, Tezpur University, Tezpur - 784028, India}

\author{Ananya Mukherjee}
\email{ananyam@tezu.ernet.in}
\affiliation{Department of Physics, Tezpur University, Tezpur - 784028, India}

\begin{abstract}
We study the possibility of generating non-zero reactor mixing angle $\theta_{13}$ and baryon asymmetry of the Universe within the framework of an $A_4$ flavour symmetric model. Using the conventional type I seesaw mechanism we construct the Dirac and Majorana mass matrices which give rise to the correct light neutrino mass matrix. Keeping the right handed neutrino mass matrix structure trivial so that it gives rise to a (quasi) degenerate spectrum of heavy neutrinos suitable for resonant leptogenesis at TeV scale, we generate the non-trivial structure of Dirac neutrino mass matrix that can lead to the light neutrino mixing through type I seesaw formula. Interestingly, such a setup naturally leads to non-zero $\theta_{13}$ due to the existence of anti-symmetric contraction of the product of two triplet representations of $A_4$.  Such antisymmetric part of triplet products usually vanish for right handed neutrino Majorana mass terms, leading to $\mu-\tau$ symmetric scenarios in the most economical setups. We constrain the model parameters from the requirement of producing the correct neutrino data as well as baryon asymmetry of the Universe for right handed neutrino mass scale around TeV. The $A_4$ symmetry is augmented by additional $Z_3 \times Z_2$ symmetry to make sure that the splitting between right handed neutrinos required for resonant leptogenesis is generated only by next to leading order terms, making it naturally small. We find that the inverted hierarchical light neutrino masses give more allowed parameter space consistent with neutrino and baryon asymmetry data.
\end{abstract}
\pacs{12.60.Fr,12.60.-i,14.60.Pq,14.60.St}
\maketitle


\section{Introduction}
Observations of tiny but non-zero neutrino mass and large leptonic mixing \cite{PDG, kamland08, T2K, chooz, daya, reno, minos} have been one of the most compelling evidences suggesting the presence of beyond standard model (BSM) physics. The present status of different neutrino parameters can be found in the latest global fit analysis \cite{schwetz16, valle17}, summarised in table \ref{tab:data1}. It can be seen that out of the three leptonic mixing angles, the solar and atmospheric angles are reasonably large while the reactor mixing angle is relatively small. On the other hand, only two mass squared differences are measured experimentally, keeping the lightest neutrino mass still an unknown parameter. Also the mass ordering is not settled yet, allowing both normal hierarchy (NH) as well as inverted hierarchy (IH). Cosmology experiments can however, put an upper bound on the lightest neutrino mass from the measurement of the sum of absolute neutrino masses $\sum_i \lvert m_i \rvert < 0.17$ eV \cite{Planck15}. Although the solar and atmospheric mixing angles $(\theta_{12}, \theta_{23})$ were known to have large values, the discovery of non-zero $\theta_{13}$ is somewhat recent \cite{T2K, chooz, daya, reno, minos}. The leptonic Dirac CP phase $\delta$ is not yet measured experimentally, though a recent measurement hinted at $\delta \approx -\pi/2$ \cite{diracphase}. If neutrinos are Majorana fermions, then two other CP phases appear, which do not affect neutrino oscillation probabilities and hence remain undetermined in such experiments. They can however be probed at experiments looking for lepton number (L) violating processes like neutrinoless double beta decay $(0\nu \beta \beta)$.
\begin{center}
\begin{table}[htb]
\begin{tabular}{|c|c|c|c|c|}
\hline
Parameters & NH \cite{schwetz16} & IH \cite{schwetz16} & NH \cite{valle17} & IH \cite{valle17} \\
\hline
$ \frac{\Delta m_{21}^2}{10^{-5} \text{eV}^2}$ & $7.03-8.09$ & $7.02-8.09 $ & $7.05-8.14$ & $7.05-8.14 $ \\
$ \frac{|\Delta m_{31}^2|}{10^{-3} \text{eV}^2}$ & $2.407-2.643$ & $2.399-2.635 $ & $2.43-2.67$ & $2.37-2.61 $ \\
$ \sin^2\theta_{12} $ &  $0.271-0.345 $ & $0.271-0.345 $ &  $0.273-0.379 $ & $0.273-0.379 $ \\
$ \sin^2\theta_{23} $ & $0.385-0.635$ &  $0.393-0.640 $ & $0.384-0.635$ &  $0.388-0.638 $ \\
$\sin^2\theta_{13} $ & $0.0193-0.0239$ & $0.0195-0.0240 $ & $0.0189-0.0239$ & $0.0189-0.0239 $ \\
$ \delta $ & $0-360^{\circ}$ & $145^{\circ}-391^{\circ}$ & $0-360^{\circ}$ & $0^{\circ}-31^{\circ}, 142^{\circ}-360^{\circ}$ \\
\hline
\end{tabular}
\caption{Global fit $3\sigma$ values of neutrino oscillation parameters \cite{schwetz16, valle17}.}
\label{tab:data1}
\end{table}
\end{center}

The standard model (SM) of particle physics, in spite of its astonishing success as a low energy theory of fundamental particles and their interactions (except gravity), can not explain the origin of neutrino mass at renormalisable level. Due to the absence of right handed neutrinos, there is no coupling of the Higgs field responsible for the origin of mass, with neutrinos. Even if right handed neutrinos are introduced, one requires a Yukawa coupling with the Higgs of the order $10^{-12}$ in order to generate sub eV neutrino masses. It also introduces a new scale, equal to the bare mass term of the right handed neutrinos that can neither be explained nor prevented within the SM. In an effective field theory setup, one can generate light neutrino masses through the dimension five effective operator \cite{weinberg} so that neutrino masses are naturally light due to the suppression by a cut-off scale $\Lambda$. Such an operator can be realised within several BSM frameworks after integrating out the heavy fields. Such renormalisable BSM frameworks are popularly known as seesaw models \cite{ti}. Apart from the tiny mass of neutrinos, another puzzling observation is their large mixing angles, in sharp contrast with small mixing angles in the quark sector. This may also be a hint that the CP violation in the leptonic sector is large compared to quark sector. If this is true, then it can have non trivial implications for cosmology as the quark sector CP violation is found to be too small to generate the observed matter antimatter asymmetry of the Universe, to be discussed in details below. The observed large mixing in the leptonic sector has motivated the study of different flavour symmetry models that can predict such mixing patterns. One of the very popular flavour symmetric scenarios is the one that predicts a $\mu-\tau$ symmetric light neutrino mass matrix that predicts $\theta_{13} = 0, \theta_{23} = \frac{\pi}{4}$ whereas the value of $\theta_{12}$ depends upon the particular realisation of this symmetry \cite{xing2015}. Among different possible realisations, the Tri-Bimaximal (TBM) \cite{Harrison} mixing pattern which predicts $\theta_{12}=35.3^o$ has probably been the most studied one. In fact, this mixing pattern was consistent with light neutrino data, prior to the discovery of non-zero $\theta_{13}$. Such mixing patterns can naturally be realised within several non-abelian discrete flavour symmetry models \cite{discreteRev}. Among them, the discrete group $A_4$ which is the group of even permutations of four objects, can reproduce the TBM mixing in the most economical way \cite{A4TBM, A4TBM1}. Since the latest neutrino oscillation data is not consistent with $\theta_{13}=0$ and hence TBM mixing, one has to go beyond the minimal $\mu-\tau$ symmetric framework. Since the measured value of $\theta_{13}$ is small compared to the other two, one can still consider the validity of $\mu-\tau$ symmetry at the leading order and generate non-zero $\theta_{13}$ by adding small $\mu-\tau$ symmetry breaking perturbations. Such corrections can originate from the charged lepton sector or the neutrino sector itself like for example, in the form of a new contribution to the neutrino mass matrix. This has led to several works including \cite{nzt13, nzt13A4, nzt13GA,db-t2, dbijmpa, dbmkdsp, dbrk, amdbmkd} within different BSM frameworks.

Apart from the issue of tiny neutrino mass and large leptonic mixing, another serious drawback of the SM is its inability to explain the observed baryon asymmetry of the Universe. The observed baryon asymmetry is often quoted as the baryon to photon ratio  \cite{Planck15}
\begin{equation}
\eta_B = \frac{n_B - n_{\overline{B}}}{n_{\gamma}} = 6.04 \pm 0.08 \times 10^{-10}
\end{equation}
If the Universe had started in a baryon symmetric manner then one has to satisfy the Sakharov's conditions \cite{sakharov}: baryon number (B) violation, C and CP violation, departure from thermal equilibrium. One popular BSM scenario that can generate a net baryon asymmetry  is leptogenesis. For a review, one may refer to \cite{davidsonPR}. As outlined in the original proposal by Fukugita and Yanagida thirty years back \cite{fukuyana}, this mechanism can satisfy all the Sakharov's conditions \cite{sakharov} required to be fulfilled in order to produce a net baryon asymmetry. Here, a net leptonic asymmetry is generated first which gets converted into baryon asymmetry through $B+L$ violating electroweak sphaleron transitions \cite{sphaleron}. The interesting feature of this scenario is that the required lepton asymmetry can be generated through out of equilibrium decay of the same heavy fields that take part in the seesaw mechanism. Although a the BSM framework explaining the baryon asymmetry could be completely decoupled from the one explaining leptonic mass and mixing, it is more economical and predictive if the same model can account for both the observed phenomena. In the conventional type I seesaw mechanism for example, the heavy right handed neutrino decay generate the required lepton asymmetry which not only depends upon the scale of right handed neutrino mass, but also on the leptonic CP violation, which can be probed at ongoing oscillation experiments. For a hierarchical spectrum of right handed neutrinos, there exists a lower bound on the right handed neutrino mass $M_R > 10^9$ GeV, popularly known as the Davidson-Ibarra bound \cite{davidsonibarra}, from the requirement of successful leptogenesis. One can however bring down the scale of right handed neutrino mass within the framework of resonant leptogenesis \cite{Pilaftsis:1997jf, Flanz:1996fb, Pilaftsis:2003gt, Xing:2006ms}.

Motivated by this, we study an $A_4$ flavour symmetric model that can simultaneously explain the correct neutrino data as well as the baryon asymmetry through TeV scale resonant leptogenesis. Keeping the right handed neutrino mass matrix trivial, giving rise to a degenerate spectrum, we first try to obtain the non-trivial Dirac neutrino mass matrix responsible for non-trivial structure of the light neutrino mass matrix, to be obtained using the type I seesaw formula. We generate this non-trivial structure of Dirac neutrino mass matrix using a flavon field which, along with the lepton doublets and right handed neutrinos transform as $A_4$ triplets. We find that this choice automatically gives rise to non-zero $\theta_{13}$ as the resulting light neutrino mass matrix do not possess any $\mu-\tau$ symmetry. This is due to the antisymmetric term arising out of the products of two $A_4$ triplets. If we generate the non-trivial leptonic mixing from a non-trivial right handed neutrino mixing, like in the Altarelli-Feruglio type models \cite{A4TBM1}, such antisymmetric term vanishes due to Majorana nature of this mass term. This is however not true in case of Dirac mass term, resulting in a non-trivial $\mu-\tau$ symmetry breaking structure in the most general case. We compare the light neutrino mass matrix derived from the model with the one from data and evaluate the model parameters for a particular choice of right handed neutrino mass scale. The minimal such scenario is found to be rather constrained with only a handful of allowed points that satisfy all the criteria from neutrino data point of view. We then feed these allowed points to the calculation of resonant leptogenesis and found agreement with the observed baryon asymmetry of the Universe. In the end we also briefly comment on the $\mu-\tau$ symmetric limit of these scenarios where the antisymmetric coupling term is turned off by hand.

This paper is organised as follows. In section \ref{sec1}, we discuss our $A_4$ flavour symmetric model with the details of different mass matrices in the lepton sector. In section \ref{sec:level1}, we briefly outline the mechanism of resonant leptogenesis followed by the details of numerical analysis in section \ref{sec:level2}. We discuss our numerical results in section \ref{sec:level3} and then briefly outline the $\mu-\tau$ symmetric limit of the model in section \ref{sec:mutau}. We finally conclude in section \ref{sec:level4}.

\section{The Model}
\label{sec1}
The discrete group $A_4$ is the group of even permutations of four objects or the symmetry group of a tetrahedron. It has twelve elements and four irreducible representations with dimensions $n_i$ such that $\sum_i n_i^2=12$. These four representations are denoted by $\bf{1}, \bf{1'}, \bf{1''}$ and $\bf{3}$ respectively. The product rules for these representations are given in appendix \ref{appen1}. We consider a flavour symmetric model based on the discrete non-abelian group $A_4$ augmented by $Z_3 \times Z_2$ which predicts the specific structures of different $3\times 3$ matrices involved in the type I seesaw in a natural and minimal way. The particle content of the model is shown in table \ref{tab1}.
\begin{table}[htb]
\begin{tabular}{|c|c|c|c|c|c|c|c|c|c|c|}
\hline  & $ \bar{L} $ & $e_R$ & $\mu_R$ & $ \tau_R $ & $N$  & $H$ & $\phi_E$ & $\phi_{\nu}$ & $\xi$ & $\zeta$ \\
\hline
$SU(2)_L$ & 2 & 1 & 1 & 1 & 1 & 2 & 1 & 1 & 1 & 1 \\
$A_4$ & 3 & 1 & $1^{\prime}$ & $1^{\prime \prime}$ & 3 & 1 & 3 & 3 & 1 & $1^{\prime \prime}$ \\
$Z_3$ & $\omega$ & $\omega^2$ & $\omega^2$ & $\omega^2$ & $\omega$ & 1 & 1 & $\omega$ & $\omega$ & 1 \\
$Z_2$ & 1 & 1 & 1 & 1 & -1 & 1 & 1 & -1 & 1 & -1 \\ 
\hline     
\end{tabular}
\caption{Fields and their transformation properties under $ SU(2)_{L} $ gauge symmetry as well as the $ A_{4} $ symmetry} \label{tab1}
  \end{table} 
The Yukawa Lagrangian for the leptons can be written as
\begin{align}
\mathcal{L}_Y & \supset Y_e \bar{L} H \frac{\phi_E}{\Lambda} e_R+Y_{\mu} \bar{L} H \frac{\phi_E}{\Lambda} \mu_R+Y_{\tau} \bar{L} H_d \frac{\phi_E}{\Lambda} \tau_R + \frac{Y_s}{\Lambda} (\phi_{\nu} \bar{L})_{3_s} \tilde{H} N + \frac{Y_a}{\Lambda} (\phi_{\nu} \bar{L})_{3_a} \tilde{H} N \nonumber \\
& +Y_N (N N)_{1} \xi + Y^{\prime}_N (N N)_{1^{\prime \prime}} \xi \frac{\zeta \zeta}{\Lambda^2} + \text{h.c.}
\end{align}
Using the $A_4$ product rules given in appendix \ref{appen1}, we can write down the relevant leptonic mass matrices from the above Lagrangian. We denote the vacuum expectation value (vev) of the Higgs to be $v_H$ and choose a specific flavon vev alignment $\langle \phi_E \rangle =(v_E, 0, 0), \langle \phi_{\nu} \rangle =(v_{\nu}, v_{\nu}, v_{\nu})$. The resulting charged lepton mass matrix is
\begin{equation}
M_l = \frac{v_H v_E}{\Lambda} \left(\begin{array}{ccc}
Y_e & 0 & 0 \\
0 & Y_{\mu} & 0 \\
0 & 0 & Y_{\tau}
\end{array}\right)
\label{cl1}
\end{equation}
The Dirac neutrino mass matrix is given by
\begin{equation}
M_D = \frac{v_H v_{\nu}}{\Lambda} \left(\begin{array}{ccc}
\frac{2}{3}Y_s & -(\frac{Y_s}{3}+\frac{Y_a}{2}) & -(\frac{Y_s}{3}-\frac{Y_a}{2}) \\
-(\frac{Y_s}{3}-\frac{Y_a}{2}) & \frac{2}{3}Y_s & -(\frac{Y_s}{3}+\frac{Y_a}{2}) \\
-(\frac{Y_s}{3}+\frac{Y_a}{2}) & -(\frac{Y_s}{3}-\frac{Y_a}{2}) & \frac{2}{3}Y_s
\end{array}\right)
\label{md1}
\end{equation}
Considering only upto dimension five terms, the right handed neutrino mass matrix can be written as
\begin{equation}
M_R =2 Y_N v_{\xi} \left(\begin{array}{ccc}
1 & 0 & 0 \\
0 & 0 & 1 \\
0 & 1 & 0 
\end{array}\right)
\label{mr1}
\end{equation}
where $ v_{\xi}$ is the vev of the flavon $\xi$. The light neutrino mass matrix can be generated using type I seesaw
\begin{equation}
-M_{\nu} = M_D M^{-1}_R M^T_D= \frac{1}{c} \left(\begin{array}{ccc}
-2 (a^2-3b^2) & (a^2+6ab-3b^2) & (a^2-6ab-3b^2) \\
(a^2+6ab-3b^2) &  (a^2-6ab-3b^2) & -2 (a^2-3b^2) \\
(a^2-6ab-3b^2) & -2 (a^2-3b^2)&  (a^2+6ab-3b^2) 
\end{array}\right)
\label{mnu1}
\end{equation}
where $a = \frac{1}{\Lambda} Y_a v_H v_{\nu}, b = \frac{2}{3 \Lambda} Y_s v_H v_{\nu}, c=2 Y_N v_{\xi}$. Diagonalisation of this mass matrix gives the eigenvalues as
\begin{equation}
m_1 = 0, \;\;\; m_2 = -\frac{3}{c}(a^2+3b^2), \;\;\; m_3 = \frac{3}{c}(a^2+3b^2)
\end{equation}
which clearly disagrees with the neutrino mass data that gives $\Delta m^2_{21} \neq 0$. Even if we lift the degeneracy of the right handed neutrino mass matrix as 
\begin{equation}
M_R =  \left(\begin{array}{ccc}
c & 0 & 0 \\
0 & 0 & c \\
0 & c & 0 
\end{array}\right) +  \left(\begin{array}{ccc}
0 & 0 & d \\
0 & d & 0 \\
d & 0 & 0 
\end{array}\right)
\end{equation}
we still have degenerate light neutrino mass eigenvalues
\begin{equation}
m_1 = 0, \;\;\; m_2 = -\frac{3(a^2+3b^2)}{\sqrt{c^2-cd+d^2}}, \;\;\; m_3 = \frac{3(a^2+3b^2)}{\sqrt{c^2-cd+d^2}}
\end{equation}
which is disallowed by neutrino data.

Choosing a more general vacuum alignment $\langle \phi_{\nu} \rangle =(v_{\nu 1}, v_{\nu 2}, v_{\nu 3})$, the Dirac neutrino mass matrix can be written as
\begin{equation}
M_D = \frac{v_H}{\Lambda} \left(\begin{array}{ccc}
\frac{2}{3}Y_s v_{\nu 1} & -(\frac{Y_s}{3}+\frac{Y_a}{2})v_{\nu 3} & -(\frac{Y_s}{3}-\frac{Y_a}{2})v_{\nu 2} \\
-(\frac{Y_s}{3}-\frac{Y_a}{2})v_{\nu 3} & \frac{2}{3}Y_s v_{\nu 2} & -(\frac{Y_s}{3}+\frac{Y_a}{2}) v_{\nu 1}\\
-(\frac{Y_s}{3}+\frac{Y_a}{2}) v_{\nu 2} & -(\frac{Y_s}{3}-\frac{Y_a}{2})v_{\nu 1} & \frac{2}{3}Y_s v_{\nu 3}
\end{array}\right)
\label{md2}
\end{equation}
Denoting $a = \frac{v_H}{\Lambda} \frac{1}{3}Y_s v_{\nu 1}, b= \frac{v_H}{\Lambda} \frac{1}{3}Y_a v_{\nu 1}, c=\frac{v_H}{\Lambda} \frac{1}{3}Y_s v_{\nu 2}, d=\frac{v_H}{\Lambda} \frac{1}{3}Y_s v_{\nu 3}$ we can write the Dirac neutrino mass matrix as 
\begin{equation}
M_D = \left(\begin{array}{ccc}
2a & -d-\frac{bd}{a} & -c+\frac{bc}{a} \\
-d+\frac{bd}{a} & 2c & -a-b \\
-c-\frac{bc}{a} & -a+b & 2d
\end{array}\right)
\label{md3}
\end{equation}
In this notation, the light neutrino mass matrix elements are given by
\begin{align}
& (-M_{\nu})_{11} =\frac{4 a^4+2 a^2 c d-2 b^2 c d}{a^2 f} \nonumber \\
&  (-M_{\nu})_{12} = \frac{a^2 (-d)+4 a b d-2 a c^2+b^2 d+2 b c^2}{a f} \nonumber \\
& (-M_{\nu})_{13} = -\frac{a^2 c+4 a b c+2 a d^2-b^2 c+2 b d^2}{a f} \nonumber \\
& (-M_{\nu})_{22} =\frac{\left(d-\frac{b d}{a}\right)^2-4 c (a+b)}{f} \nonumber \\
& (-M_{\nu})_{23} = \frac{a^4-a^2 \left(b^2-5 c d\right)-b^2 c d}{a^2 f} \nonumber \\
& (-M_{\nu})_{33} = \frac{\frac{c^2 (a+b)^2}{a^2}+4 d (b-a)}{f}
\label{mnu2}
\end{align}
where $f=2 Y_N v_{\xi}$ is the non-zero entry in the right handed neutrino mass matrix given by \eqref{mr1}.
In this case, the resulting light neutrino mass matrix can give rise to the correct mass squared differences as well as mixing angles including non-zero $\theta_{13}$. At the dimension five level however, the right handed neutrinos remain degenerate. As we discuss below, for successful resonant leptogenesis, the right handed neutrinos must have tiny splittings which can be generated at dimension six level in the model. This higher order contribution to the right handed neutrino mass matrix can be written as
\begin{equation}
\delta M = \left(\begin{array}{ccc}
      0 & 0 & r_1\\
      0 & r_1 & 0 \\ 
      r_1 & 0 & 0
      \end{array}\right)
      \end{equation}
where $r_1 = Y^{\prime}_N v_{\xi} \frac{v^2_{\zeta}}{\Lambda^2}$ with $v_{\zeta}$ being the vev of the flavon $\zeta$. Such a small higher order term does not affect light neutrino masses and mixings considerably.

It should be noted that, we have used the $A_4$ product rules in $T$ diagonal basis, as given in appendix \ref{appen1}. This is justified in the diagonal charged lepton and Majorana light neutrino mass limit. In the $S$ diagonal basis, the charged lepton mass matrix is non-diagonal and the light neutrino mass matrix will also have a different structure due to the difference in the triple product rules. For details, one may refer to \cite{discreteRev}.
\begin{figure*}[h!]
\begin{center}
\includegraphics[width=15cm,height=3cm]{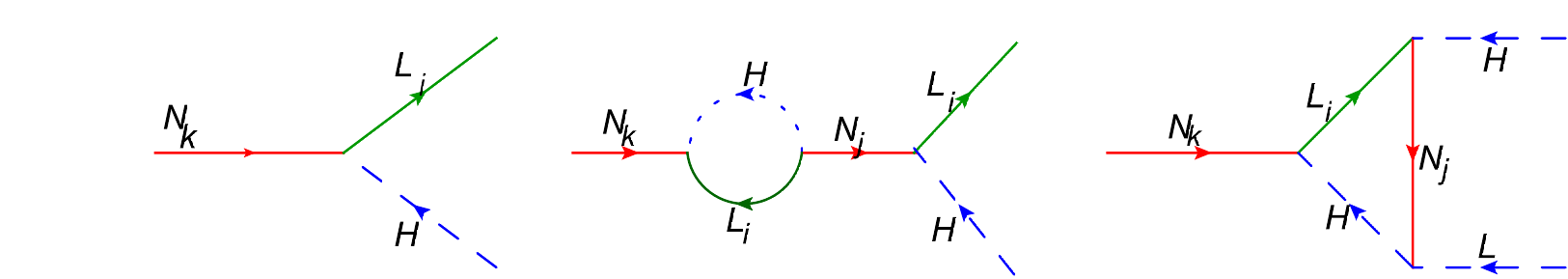}
\caption{Decay modes of right handed neutrino in type I seesaw}
\label{fig1}
\end{center}
\end{figure*}
\section{\label{sec:level1}Resonant Leptogenesis}
As pointed out by Fukugita and Yanagida \cite{fukuyana}, the out of equilibrium and CP violating decay of heavy Majorana neutrinos provides a natural way to create the required lepton asymmetry, as shown in figure \ref{fig1}. The asymmetry generated by the decay of the lightest right handed neutrino into lepton and Higgs is given by, 
    \begin{equation}
    \epsilon_{N_{k}} = -\sum \frac{\Gamma(N_{k} \rightarrow L_{i}+H^{*})-\Gamma(N_{k} \rightarrow L_{i}+H)}
    {\Gamma(N_{k} \rightarrow L_{i}+H^{*})+\Gamma(N_{k} \rightarrow L_{i}+H)}
   \end{equation} 
This lepton asymmetry is converted to the baryon asymmetry through electroweak sphaleron transitions allowing us to reproduce the observed baryon asymmetry of the Universe. As mentioned before, resonant leptogenesis is a viable alternative to high scale or vanilla leptogenesis scenarios \cite{Pilaftsis:1997jf,Flanz:1996fb,Pilaftsis:2003gt,Xing:2006ms} within the context of a TeV scale minimal seesaw scenarios. Since a hierarchical spectrum of right handed neutrinos can not give rise to the required asymmetry at TeV scale, this mechanism gives a resonance enhancement to the lepton asymmetry by considering a very small mass splitting between the two heavy neutrinos, of the order of their average decay width.

The lepton asymmetry can be found from the following formula \cite{dev2014, Bambhaniya:2016rbb},
 \begin{equation}\label{eq:asymmetry}
    \epsilon_{il} = \sum_{j \neq i} \frac{\text{Im}[Y_{\nu_{il}}Y_{\nu_{jl}}^{*}(Y_{\nu}Y_{\nu}^\dagger)_{ij}]+\frac{M_i}{M_j}\text{Im}[Y_{\nu_{il}} Y_{\nu_{jl}}^{*}(Y_{\nu}Y_{\nu}^\dagger)_{ji}]}{(Y_{\nu}Y_{\nu}^{\dagger})_{ii}(Y_{\nu}Y_{\nu}^{\dagger})_{jj}} f_{ij}  
   \end{equation}
with the regulator $f_{ij}$ being given as
\begin{equation*}
f_{ij} = \frac{(M_{i}^2 - M_j^2)M_{i}\Gamma_{j}}{(M_{i}^2 - M_j^2)^2 + M_i^2 \Gamma_j^2}.  
\end{equation*}
Here, $\Gamma_i = \frac{M_i}{8\pi}(Y_\nu Y_\nu^\dagger)_{ii}$ as the tree level heavy-neutrino decay width and $Y_{\nu}$ is the effective coupling between heavy and light neutrinos. Now, there is a similar contribution $\epsilon_{il}^{\prime}$ to the CP asymmetry from RH neutrino oscillation \cite{Kartavtsev:2015vto, dev2014, dev1410}. Its form is given by \eqref{eq:asymmetry} with the replacement $f_{ij}$ by $f_{ij}^{\prime}$, where 
\begin{equation*}
f_{ij}^{\prime} = \frac{(M_{i}^2 - M_j^2)M_{i}\Gamma_{j}}{(M_{i}^2 - M_j^2)^2 + (M_i \Gamma_i +M_j \Gamma_j)^2 \frac{\text{det}[\text{Re}(Y_\nu Y_\nu^\dagger)]}{(Y_\nu Y_\nu^\dagger)_{ii}(Y_\nu Y_\nu^\dagger)_{ii}}}.  
\end{equation*}
The total CP asymmetry is therefore is the summation of these two $\epsilon^T_{il} = \epsilon_{il}+\epsilon^{\prime}_{il}$. Taking into account of the appropriate efficiency and dilution factors \cite{dev2014, Deppisch:2010fr}, one can write the final baryon asymmetry as
 \begin{equation} \label{eq:bau}
 \eta_B =\frac{n_B - n_{\overline{B}}}{n_{\gamma}} \simeq -\frac{28}{51} \frac{1}{27} \frac{3}{2} \sum_{l,i}\frac{\epsilon_{il}}{K_l^{\text{eff}}\text{min}(z_c,z_l)}
\end{equation}  
where, $z_c = \frac{M_N}{T_c}$, $T_c \sim 149$ GeV being the critical temperature, $z_l \simeq 1.25 \text{log}(25 K_l^{\text{eff}})$ \cite{Deppisch:2010fr} and $K_l^{\text{eff}} = \kappa_l \sum_{i} K_i B_{il}$, with $K_i = \Gamma_i /H_N$ being the wash out factor. The Hubble parameter for radiation dominated Universe is $H_N=1.66 \sqrt{g_*}M_N^2/M_{\text{Pl}}$ at $T=M_N$ and $g^* \simeq 106.75$ is the relativistic degrees of freedom at high temperatures. $B_{il}$'s are the branching ratios of the $N_i$ decay to leptons of lth flavor: $B_{il} = \frac{|Y_{\nu_{il}}|^2}{(Y_{\nu}Y_{\nu}^{\dagger})_{ii}}$. The factor $\kappa$ is given by
\begin{equation}
\kappa_l = 2 \sum_{i,j (j \neq i)} \frac{\text{Re}[Y_{\nu il}Y_{\nu jl}^* (YY^\dagger)_{ij}]+ \text{Im}[(Y_{\nu il} Y_{\nu jl}^*)^2]}{\text{Re}[(Y^\dagger Y)_{ll} \{(Y Y^\dagger)_{ii} + (Y Y^\dagger)_{jj}\}]} \left (1-2i \frac{M_i-M_j}{\Gamma_i + \Gamma_j} \right)^{-1}.
\end{equation} 
As seen from the expression \eqref{eq:asymmetry}, the lepton asymmetry is dependent on the elements of the Dirac Yukawa coupling matrix. Therefore it can be said that, the same sets of model parameters which are supposed to yield correct neutrino phenomenology are also responsible to yield an enhanced lepton asymmetry, later on generating the observed BAU.

 \section{\label{sec:level2} Numerical Analysis}
 As discussed before, the most general form of Dirac neutrino mass matrix (assuming a degenerate right handed neutrino mass spectrum) can give rise to a light neutrino mass matrix from type I seesaw formula, which is consistent with $\theta_{13} \neq 0$. This is due to the presence of anti-symmetric part of $A_4$ triple product that explicitly breaks $\mu-\tau$ symmetry leading to the generation of $\theta_{13} \neq 0$. Within the minimal setup, the light neutrino mass matrix is given by \eqref{mnu2}, which contains five parameters $a, b, c, d, f$ that can in general be complex. Since this corresponds to degenerate heavy neutrino masses which can not give rise to successful leptogenesis, we can break the degeneracy by including higher order contribution to the right handed neutrino mass matrix as discussed above. Taking this correction into account, we can write the right handed neutrino mass matrix as
  \begin{equation}\label{eq:M}
      M = M^{0}_R + \delta M_R = \left(\begin{array}{ccc}
      f & 0 & g\\
      0 & g & f \\ 
      g & f & 0
      \end{array}\right)
      \end{equation}
This has eigenvalues $f + g, -\sqrt{f^2 - f g + g^2}, \sqrt{f^2 - f g + g^2}$ where, $f$ is the leading order right handed neutrino mass and $g$ is the parameter creating tiny mass splitting.  
As mentioned earlier, these parameters are related to the Lagrangian parameters as
$$a = \frac{v_H}{\Lambda} \frac{1}{3}Y_s v_{\nu 1}, b= \frac{v_H}{\Lambda} \frac{1}{3}Y_a v_{\nu 1}, c=\frac{v_H}{\Lambda} \frac{1}{3}Y_s v_{\nu 2}, d=\frac{v_H}{\Lambda} \frac{1}{3}Y_s v_{\nu 3}, f=2 Y_N v_{\xi}, g=Y^{\prime}_N v_{\xi} \frac{v^2_{\zeta}}{\Lambda^2}$$

For numerical analysis part we first fix the scale of leptogenesis by fixing the leading right handed neutrino mass or the parameter $f$ to be 5 TeV, say. The range of $g$ has been chosen in such a way that we can have a tiny Majorana mass splitting required for successful leptogenesis without affecting the neutrino parameters being from their correct $3\sigma$ bound. For satisfying neutrino phenomenology and explaining leptogenesis, $g$ has been varied randomly from $10^{-6}$ to $10^{-5}$ GeV which gives lepton asymmetry of an order around $10^{-7}$ or more. Since $g$ is very small compared to $f$, its effects on light neutrino masses and mixing is not substantial. Yet, we include it while discussing the compatibility of the model with neutrino data. Thus, after making the choice of $f$ and the range of $g$, we are left with four model parameters $a, b, c, d$ that can be calculated by comparing the mass matrix predicted by the model with the one we can construct in terms of light neutrino parameters.

The leptonic mixing matrix can be written in terms of the charged lepton diagonalising matrix $(U_l)$ and light neutrino diagonalising matrix $U_{\nu}$ as 
\begin{equation}
U_{\text{PMNS}} = U^{\dagger}_l U_{\nu}
\end{equation}
In the simple case where the charged lepton mass matrix is diagonal which is true in our model, we can have $U_l = \mathbb{1}$. Therefore we can write $U_{\text{PMNS}} = U_{\nu}$. Now we can write the complete light neutrino mass matrix as 
\begin{equation}
m_{\nu}= U_{\text{PMNS}}m^{\text{diag}}_{\nu} U^T_{\text{PMNS}}
\label{nuoscillation}
\end{equation}
where the Pontecorvo-Maki-Nakagawa-Sakata (PMNS) leptonic mixing matrix can be parametrised as
\begin{equation}
U_{\text{PMNS}}=\left(\begin{array}{ccc}
c_{12}c_{13}& s_{12}c_{13}& s_{13}e^{-i\delta}\\
-s_{12}c_{23}-c_{12}s_{23}s_{13}e^{i\delta}& c_{12}c_{23}-s_{12}s_{23}s_{13}e^{i\delta} & s_{23}c_{13} \\
s_{12}s_{23}-c_{12}c_{23}s_{13}e^{i\delta} & -c_{12}s_{23}-s_{12}c_{23}s_{13}e^{i\delta}& c_{23}c_{13}
\end{array}\right) U_{\text{Maj}}
\label{PMNS}
\end{equation}
where $c_{ij} = \cos{\theta_{ij}}, \; s_{ij} = \sin{\theta_{ij}}$ and $\delta$ is the leptonic Dirac CP phase. The diagonal matrix $U_{\text{Maj}}=\text{diag}(1, e^{i\alpha}, e^{i(\zeta+\delta)})$ contains the undetermined Majorana CP phases $\alpha, \zeta$. The diagonal mass matrix of the light neutrinos can be written  as $m^{\text{diag}}_{\nu} 
= \text{diag}(m_1, \sqrt{m^2_1+\Delta m_{21}^2},\sqrt{m_1^2+\Delta m_{31}^2})$ for normal hierarchy (NH) and $m^{\text{diag}}_{\nu} = \text{diag}(\sqrt{m_3^2+\Delta m_{23}^2-\Delta m_{21}^2}$, $\sqrt{m_3^2+\Delta m_{23}^2}, m_3)$ for inverted hierarchy (IH). 

For a fixed value of right handed neutrino mass, we can now compare the light neutrino mass matrix predicted by the model and the one calculated from the light neutrino parameters. Since there are four undetermined complex parameters of the model, we need to compare four elements. Without any loss of generality, we equate $(12), (13), (22), (33)$ elements of both the mass matrices. We vary the light neutrino parameters in their allowed $3\sigma$ ranges and vary the lightest neutrino mass $m_{\text{lightest}} \in (10^{-6}, 0.1) $ eV and calculate the model parameters $a, b, c, d$ for each set of values of neutrino parameters. However, the light neutrino mass matrix has two more independent elements as any general $3\times 3$ complex symmetric mass matrix has six independent complex elements. On the other hand, once $a, b, c, d$ are calculated from the equations $(M_{\nu})_{12}=(m_{\nu})_{12}, (M_{\nu})_{13}=(m_{\nu})_{13}, (M_{\nu})_{22}=(m_{\nu})_{22}, (M_{\nu})_{33}=(m_{\nu})_{33}$, the other two elements $(M_{\nu})_{11}, (M_{\nu})_{23}$ are automatically determined. Since every set of values of $a, b, c, d$ corresponds to a particular set of light neutrino parameters, we can calculate the other two light neutrino mass matrix elements $(m_{\nu})_{11}, (m_{\nu})_{23}$ for the same set of neutrino parameters. For consistency, one needs to make sure that these two elements calculated for the neutrino mass matrix predicted by the model $M_{\nu}$ and the ones from light neutrino parameters $m_{\nu}$ are equal to each other. It turns out that these two constraints tightly restrict the light neutrino parameters to a set of very specific values, resulting in a very predictive scenario. We randomly generate ten million light neutrino parameters to calculate the four model parameters $a, b, c, d$ and restrict the parameters to only those ones which satisfy $\lvert (m_{\nu})_{11}-(M_{\nu})_{11} \rvert < 10^{-5}, \lvert (m_{\nu})_{23}-(M_{\nu})_{23} \rvert < 10^{-5}$. Here a tolerance of $10^{-5}$ is chosen to decide the equality between the two elements.

After finding the model parameters $a, b, c, d$ as well as the light neutrino parameters satisfying the constraints relating the two elements of the mass matrices constructed from the model and neutrino data respectively, we calculate the lepton asymmetry for the same set of allowed parameters. The effective Dirac Yukawa coupling matrix ($Y_\nu$) relating heavy neutrinos to the light ones appearing in the lepton asymmetry formula is considered to have the same structure as the Dirac neutrino mass matrix given in \eqref{md2}. Since the corrected form of the heavy neutrino mass matrix is non-diagonal (given by \eqref{eq:M}), we first diagonalise it and find the corresponding diagonalising matrix $U_R$. To keep the analysis in this basis we transform the Dirac Yukawa coupling matrices as $Y_{\nu} \rightarrow Y_{\nu} U_R$ with $U_R^* M_R U_R^\dagger = \text{diag}(M_1, M_2, M_3)$. We then calculate the baryon asymmetry for the light neutrino parameters that are consistent with neutrino data as well as the model restrictions discussed above.

\begin{figure*}
\begin{center}
\includegraphics[width=0.45\textwidth]{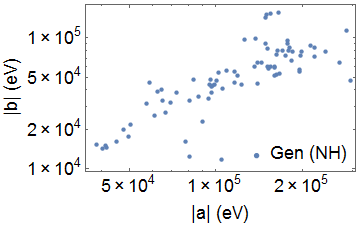}
\includegraphics[width=0.45\textwidth]{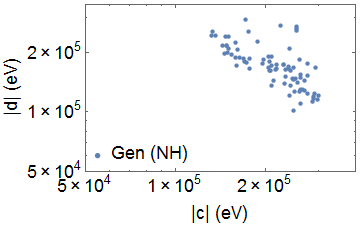}
\end{center}
\begin{center}
\caption{Correlation between different model parameters for normal hierarchy. The label Gen refers to the most general structure of the mass matrix discussed in the text.}
\label{fig2}
\end{center}
\end{figure*}
 \begin{figure*}
\begin{center}
\includegraphics[width=0.45\textwidth]{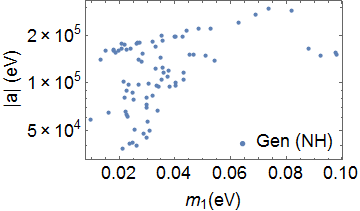}
\includegraphics[width=0.45\textwidth]{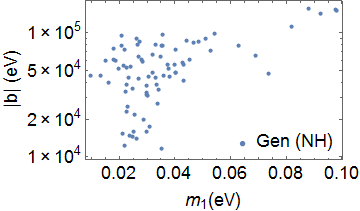} \\
\includegraphics[width=0.45\textwidth]{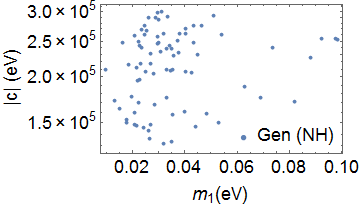}
\includegraphics[width=0.45\textwidth]{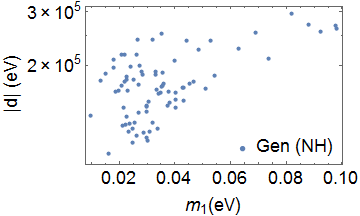}
\end{center}
\begin{center}
\caption{Model parameters as a function of the lightest neutrino mass for normal hierarchy. The label Gen refers to the most general structure of the mass matrix discussed in the text.}
\label{fig3}
\end{center}
\end{figure*}
 \begin{figure*}
\begin{center}
\includegraphics[width=0.45\textwidth]{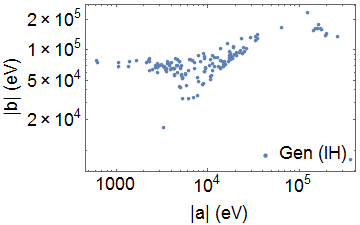}
\includegraphics[width=0.45\textwidth]{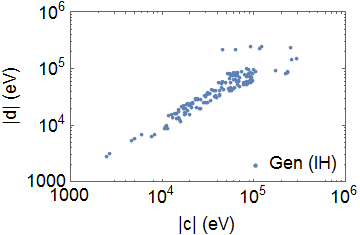} 
\end{center}
\begin{center}
\caption{Correlation between different model parameters for inverted hierarchy. The label Gen refers to the most general structure of the mass matrix discussed in the text.}
\label{fig4}
\end{center}
\end{figure*}
 \begin{figure*}
\begin{center}
\includegraphics[width=0.45\textwidth]{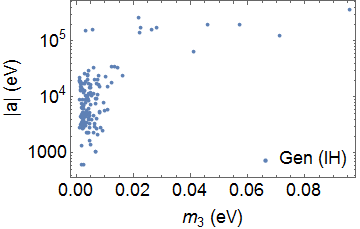}
\includegraphics[width=0.45\textwidth]{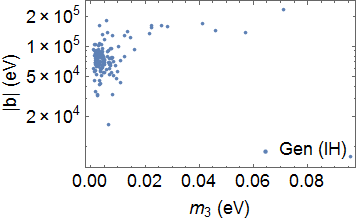} \\
\includegraphics[width=0.45\textwidth]{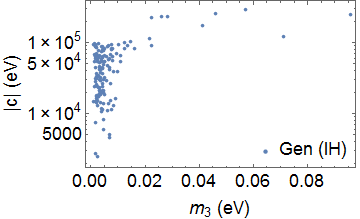}
\includegraphics[width=0.45\textwidth]{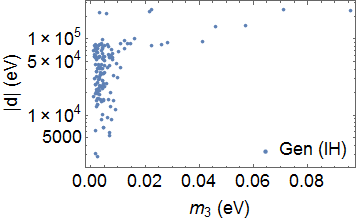}
\end{center}
\begin{center}
\caption{Model parameters as a function of the lightest neutrino mass for inverted hierarchy. The label Gen refers to the most general structure of the mass matrix discussed in the text.}
\label{fig5}
\end{center}
\end{figure*}
\begin{figure*}
\begin{center}
\includegraphics[width=0.45\textwidth]{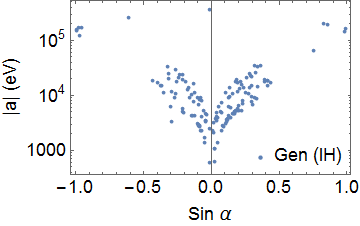}
\includegraphics[width=0.45\textwidth]{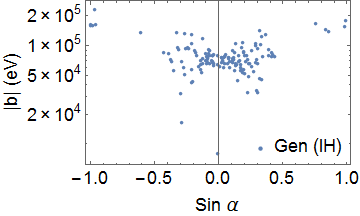} \\
\includegraphics[width=0.45\textwidth]{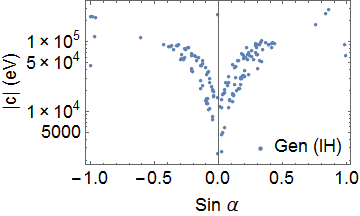}
\includegraphics[width=0.45\textwidth]{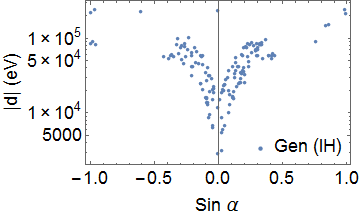}
\end{center}
\begin{center}
\caption{Model parameters as a function of one of the Majorana phases $\alpha$ for inverted hierarchy. The label Gen refers to the most general structure of the mass matrix discussed in the text.}
\label{fig6}
\end{center}
\end{figure*}

\begin{figure*}
\begin{center}
\includegraphics[width=0.45\textwidth]{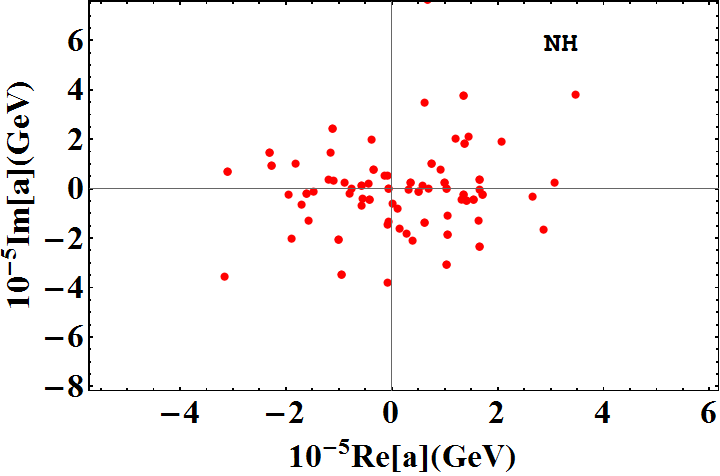}
\includegraphics[width=0.45\textwidth]{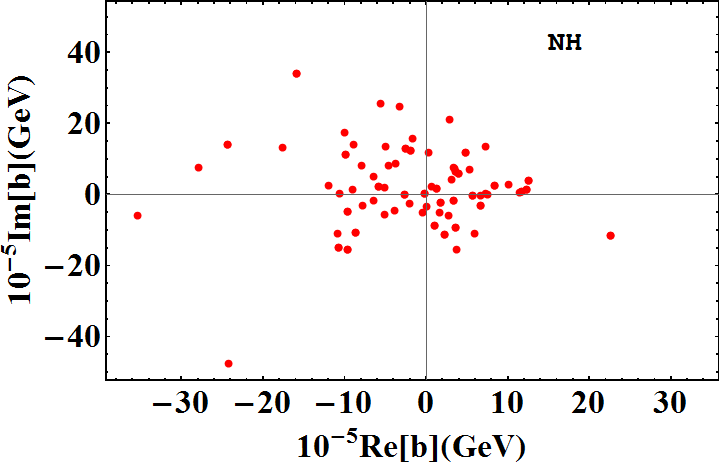} \\
\includegraphics[width=0.45\textwidth]{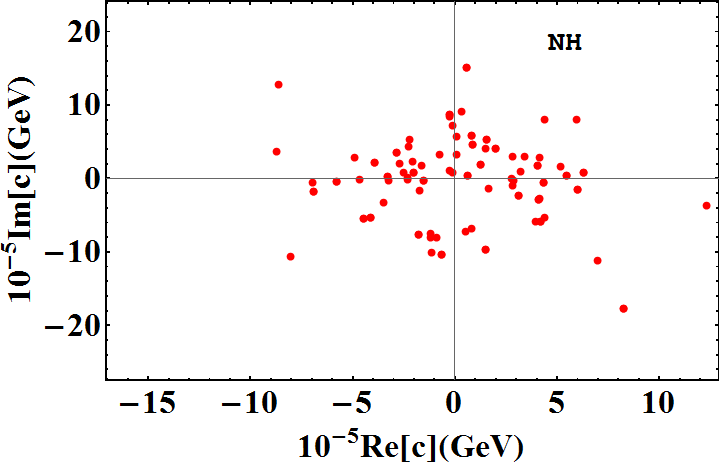}
\includegraphics[width=0.45\textwidth]{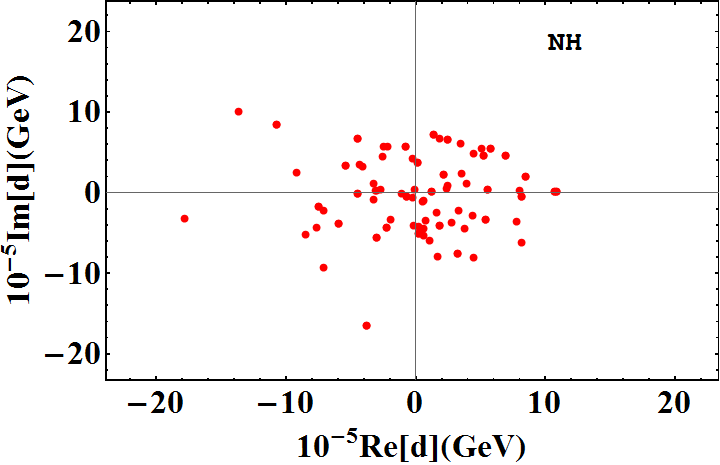}
\end{center}
\begin{center}
\caption{Real and imaginary parts of the model parameters for normal hierarchy with the most general structure of the mass matrix discussed in the text.}
\label{fig6a}
\end{center}
\end{figure*}

\begin{figure*}
\begin{center}
\includegraphics[width=0.45\textwidth]{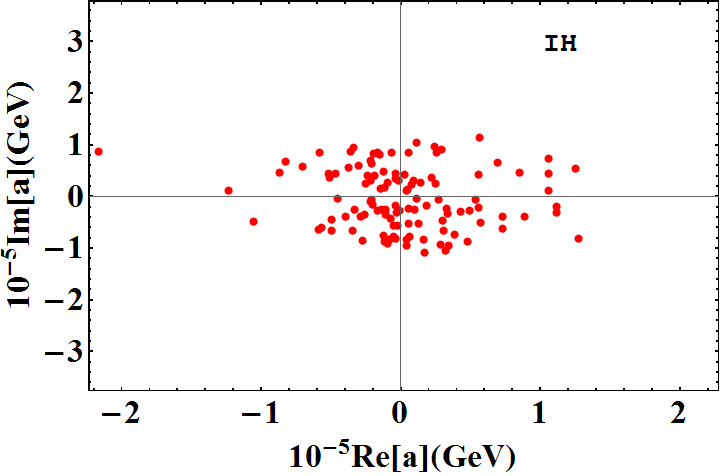}
\includegraphics[width=0.45\textwidth]{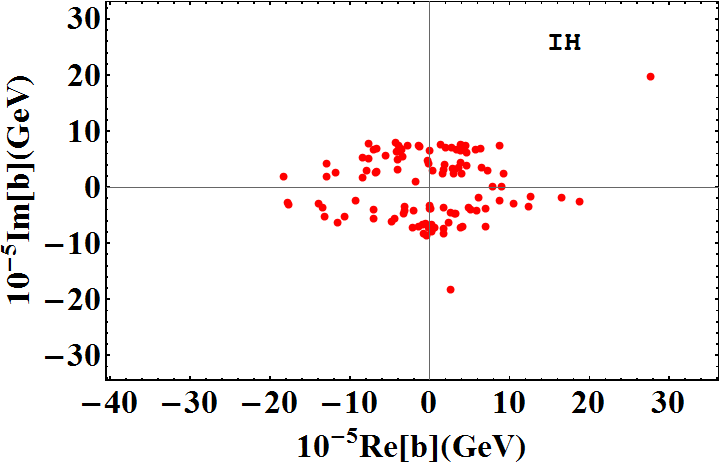} \\
\includegraphics[width=0.45\textwidth]{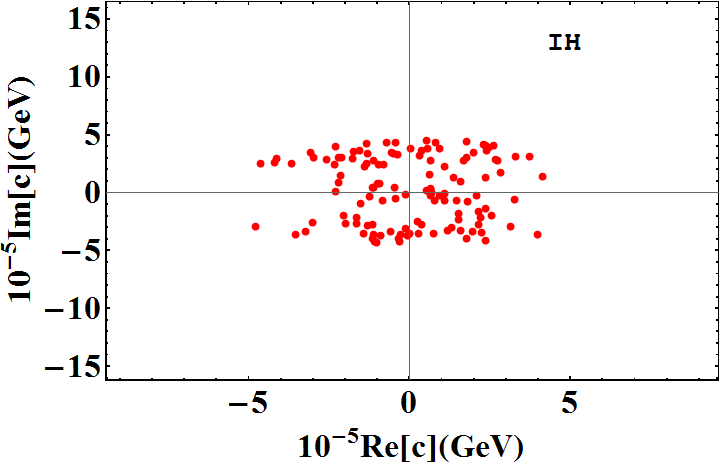}
\includegraphics[width=0.45\textwidth]{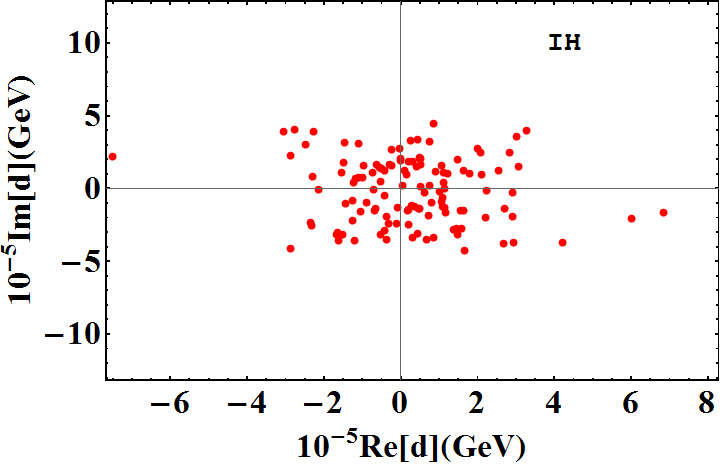}
\end{center}
\begin{center}
\caption{Real and imaginary parts of the model parameters for inverted hierarchy with the most general structure of the mass matrix discussed in the text.}
\label{fig6b}
\end{center}
\end{figure*}

\begin{figure*}
\begin{center}
\includegraphics[width=0.45\textwidth]{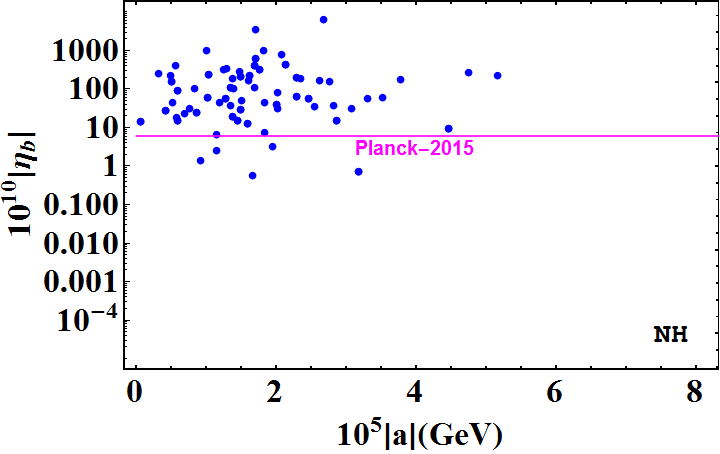}
\includegraphics[width=0.45\textwidth]{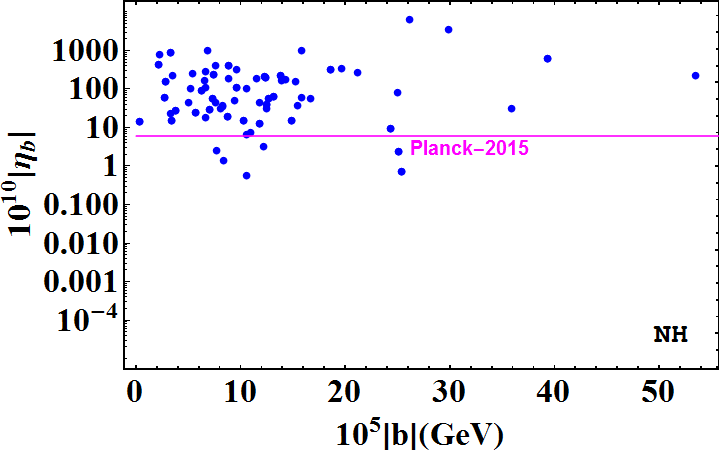} \\
\includegraphics[width=0.45\textwidth]{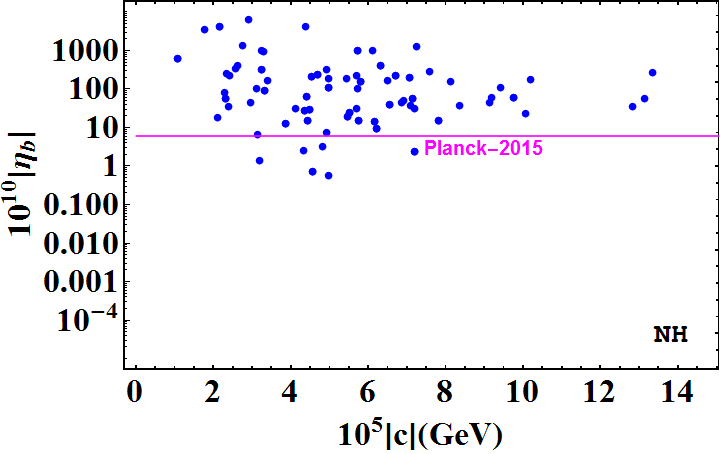}
\includegraphics[width=0.45\textwidth]{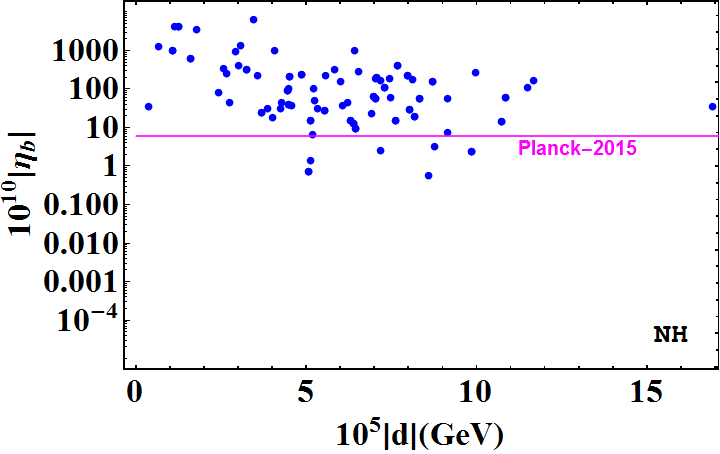}
\end{center}
\begin{center}
\caption{Baryon asymmetry as a function of model parameters for normal hierarchy. The horizontal pink line corresponds to the Planck bound $\eta_B = 6.04 \pm 0.08 \times 10^{-10}$ \cite{Planck15}.}
\label{fig7}
\end{center}
\end{figure*}

\begin{figure*}
\begin{center}
\includegraphics[width=0.45\textwidth]{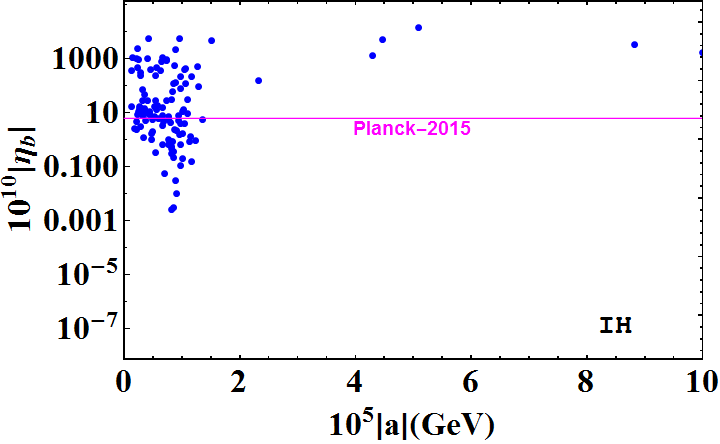}
\includegraphics[width=0.45\textwidth]{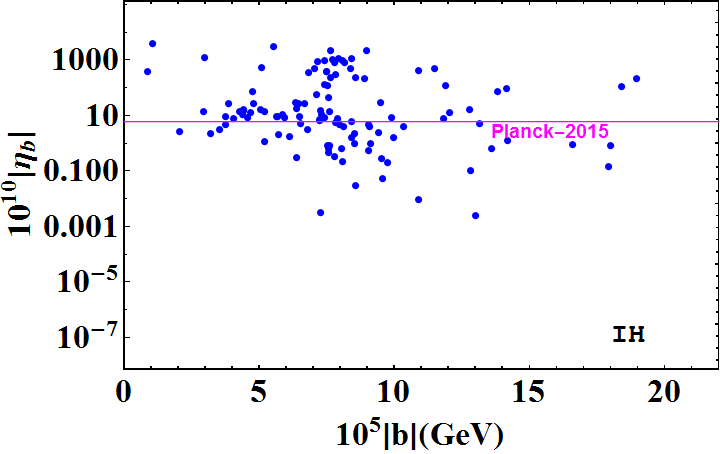} \\
\includegraphics[width=0.45\textwidth]{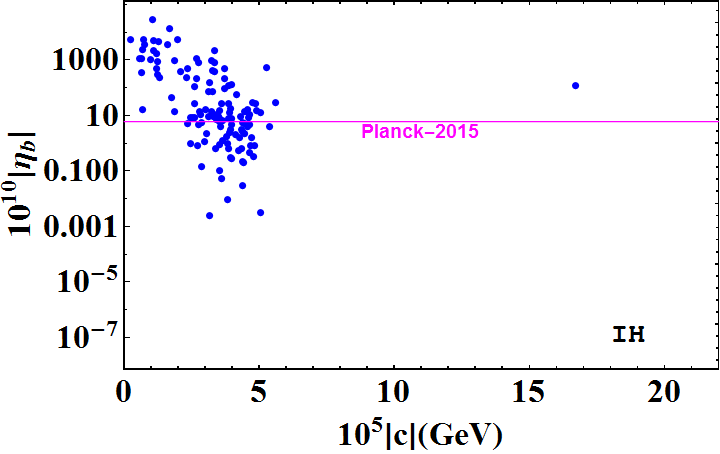}
\includegraphics[width=0.45\textwidth]{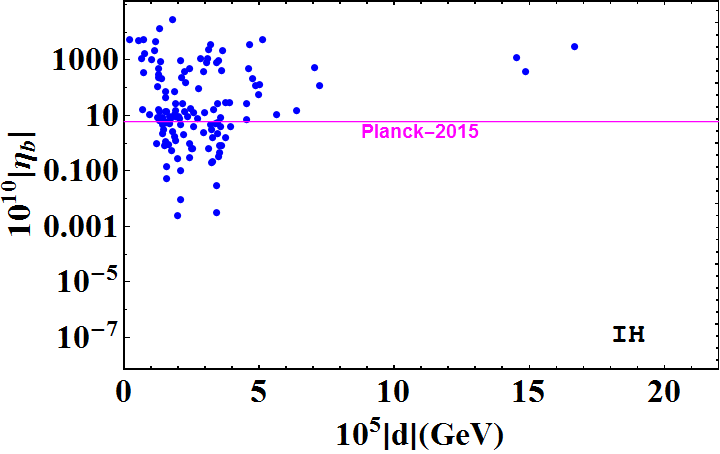}
\end{center}
\begin{center}
\caption{Baryon asymmetry as a function of model parameters for inverted hierarchy. The horizontal pink line corresponds to the Planck bound $\eta_B = 6.04 \pm 0.08 \times 10^{-10}$ \cite{Planck15}.}
\label{fig8}
\end{center}
\end{figure*}

\begin{figure*}
\begin{center}
\includegraphics[width=0.45\textwidth]{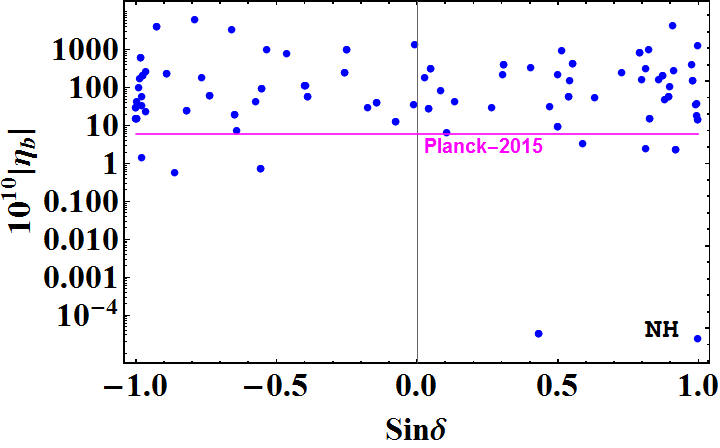}
\includegraphics[width=0.45\textwidth]{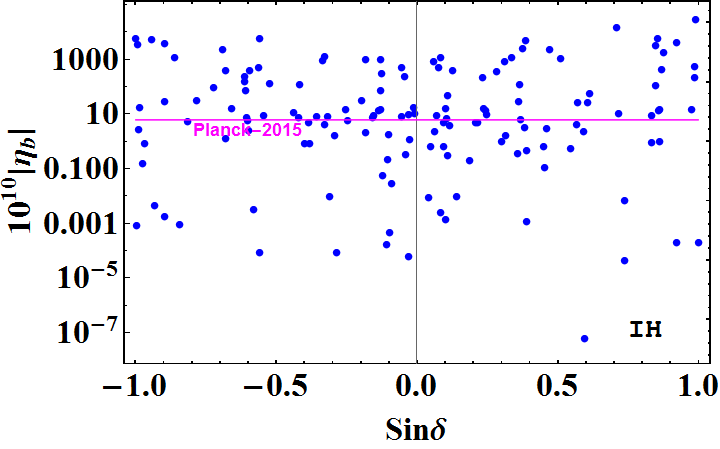} 
\end{center}
\begin{center}
\caption{Baryon asymmetry as a function of Dirac CP phase for normal and inverted hierarchy. The horizontal pink line corresponds to the Planck bound $\eta_B = 6.04 \pm 0.08 \times 10^{-10}$ \cite{Planck15}.}
\label{fig9}
\end{center}
\end{figure*}

\begin{figure*}
\begin{center}
\includegraphics[width=0.45\textwidth]{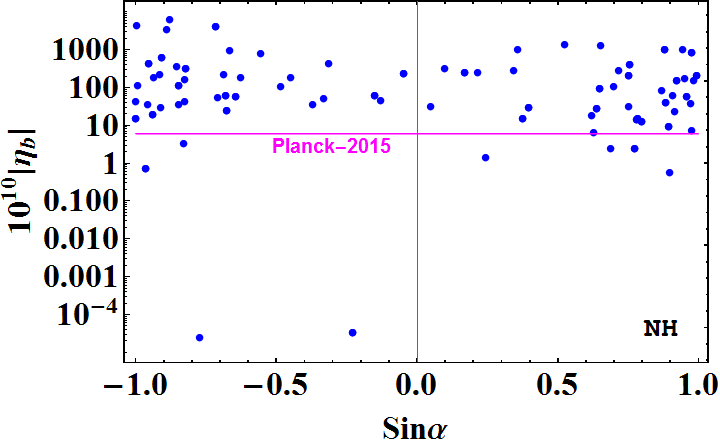}
\includegraphics[width=0.45\textwidth]{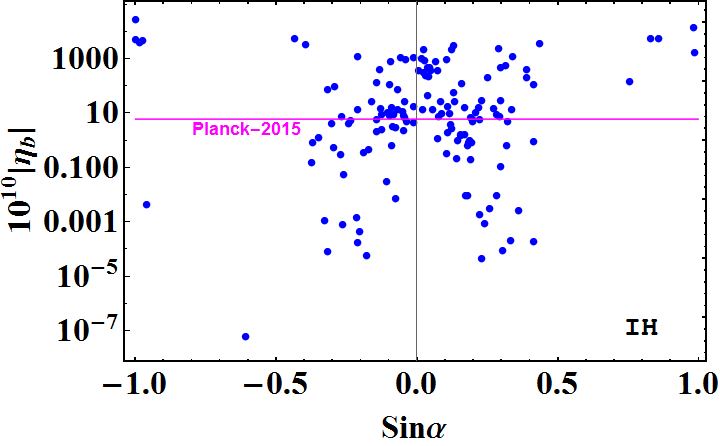} \\
\includegraphics[width=0.45\textwidth]{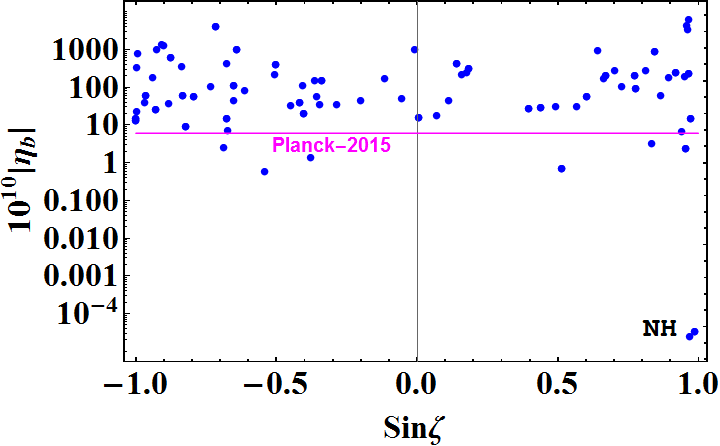}
\includegraphics[width=0.45\textwidth]{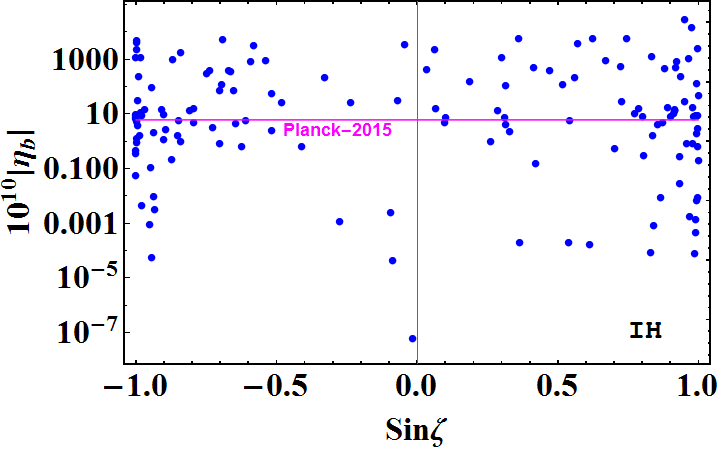}
\end{center}
\begin{center}
\caption{Baryon asymmetry as a function of Majorana CP phases for normal and inverted hierarchy. The horizontal pink line corresponds to the Planck bound $\eta_B = 6.04 \pm 0.08 \times 10^{-10}$ \cite{Planck15}.}
\label{fig10}
\end{center}
\end{figure*}

\section{\label{sec:level3}Results and Discussion}
Following the procedures outlined in the previous section, we first randomly generate the light neutrino parameters in their $3\sigma$ range \cite{schwetz16} and for each set of values, we calculate the model parameters $a, b, c, d$ using four equations. We then apply the constraints relating other two elements of the neutrino mass matrix and find the constrained parameter space obeying them. For normal hierarchy, we show the correlation between these model parameters in figure \ref{fig2}. Since $a, b, c, d$ denote the strength of the Dirac neutrino mass, we can see that they lie near or below the MeV scale so that the correct light neutrino mass is generated from type I seesaw formula where the right handed neutrino scale is fixed at 5 TeV. We also show the variation of the same model parameters with the lightest neutrino mass $m_1$ for normal hierarchy in figure \ref{fig3}. It can be seen that the allowed lightest neutrino mass can have values in the range $0.01-0.1$ eV, that can be sensitive to $0\nu \beta \beta$ experiments. In fact, the region of parameter space near $m_1 \sim 0.1$ eV will be ruled out by latest bounds from $0\nu \beta \beta$ experiments as well as the cosmology upper bound on the sum of absolute neutrino masses \cite{Planck15}. We show similar correlations for inverted hierarchy in figure \ref{fig4}, \ref{fig5}. The overfall features of these correlation plots are similar the ones for normal hierarchy, shown in figure \ref{fig2}, \ref{fig3}. However, for inverted hierarchy, we see a preference for smaller values of lightest neutrino mass, close to 0.01 eV, away from the upper bounds set by $0\nu \beta \beta$ and cosmology data. We then show some interesting correlations between the model parameters for inverted hierarchy with one of the Majorana CP phases in figure \ref{fig6}. This figure also shows that the requirement of satisfying correct neutrino data constrains this CP phase to a range $\lvert \sin{\alpha} \rvert < 0.5$.
 
 We also check if there are any correlations among the known neutrino parameters in this analysis. This could arise due to the fact that there are only four parameters $a, b, c, d$ that we are solving for by using more number of input parameters, leading to an over-constrained system. However, we did not find any such correlations between the known neutrino parameters. This is primarily due to the fact that the model parameters $a, b, c, d$ are in general complex and hence they represent a set of eight real parameters. We show their real and imaginary parts separately in figure \ref{fig6a} and \ref{fig6b} for normal and inverted hierarchies respectively. The imaginary parts of the model parameters are the source of CP phases in this model and hence play a crucial role in generating the leptonic asymmetries.

After finding the allowed neutrino as well as model parameters from the requirement of satisfying the latest neutrino oscillation data, we feed them to the calculation of the baryon asymmetry through resonant leptogenesis. The resulting values of $\eta_B$ are shown for normal hierarchy as a function of the model parameters in figure \ref{fig7}. We can see that there are several points which satisfy the Planck 2015 bound on baryon asymmetry $\eta_B = 6.04 \pm 0.08 \times 10^{-10}$ \cite{Planck15}. We find more allowed parameters that satisfy the Planck bound for inverted hierarchy, as can be seen from the plots shown in figure \ref{fig8}. We also show the baryon asymmetry versus Dirac CP phase $\delta$ in figure \ref{fig9}. It can be seen from this plot that, we do not see preference for any particular value of Dirac CP phase. To show the variation of $\eta_B$ with Majorana CP phases, we show the plots in figure \ref{fig10} for both normal and inverted hierarchy.

 Here we note that there is a difference of around nine order of magnitudes between the mass splitting between the right handed neutrinos (of keV order) and their masses (of TeV order). Although in this model we have generated such tiny mass splitting naturally, by forbidding it at leading order and generating it only at higher orders (mass splitting term is suppressed by $\Lambda^2$ compared to the dimension four mass term without any suppression, as discussed above), we still need to make sure that these splittings are stable under quantum corrections. That is, if we generate this tiny splitting naturally at the scale of the flavour symmetry breaking $\sim \Lambda$, such splittings should not be disturbed significantly while running them down to the scale at which the lepton asymmetry is being generated $T \sim M_R \sim \mathcal{O}(\text{TeV})$. Several earlier works discussed such radiative origin of mass splittings \cite{radiative1} by considering a degenerate spectrum at high energy scale \cite{dev1410, radiative2}. Such splittings  at the scale of leptogenesis ($T \sim M_R$) originating from renormalisation group (RG) effects from a scale $\Lambda$ to $M_R$ can be estimated as
\begin{equation}
\Delta M^{\text{RG}}_R \approx -\frac{M_R}{8\pi^2} \ln{\left( \frac{\Lambda}{M_R} \right)} \text{Re}[ Y^{\dagger}_{\nu} (\Lambda) Y_{\nu}(\Lambda)] 
 \end{equation}
 The effective Yukawa couplings $Y_{\nu}$ here can be derived from the model parameters $a, b, c, d$ by taking their ratio with the Higgs vev $v_H \sim 100$ GeV. As seen from the figures \ref{fig6a}, \ref{fig6b}, the parameters $a, b, c, d$ can be as large as of order $10^{-4}$ GeV and hence the effective Yukawa couplings $Y_{\nu}$ will be of the order of $10^{-6}$. Thus, the mass splitting from RG effects can be estimated to be approximately 
 $$ \Delta M^{\text{RG}}_R \approx (x-3) \times 3 \times 10^{-11} \; \text{GeV}$$
 where $\Lambda =10^x$ GeV, $M_R \sim \mathcal{O}(\text{TeV})$ is used. Therefore, the splitting from RG effects is usually small for TeV scale $M_R$ and the values of Yukawa couplings we have in our model. In fact, as pointed out by \cite{Dev:2015wpa}, pure radiative splitting scenario gives rise to vanishing lepton asymmetry at order $\mathcal{O}(Y^4_{\nu})$, showing more preference to non-minimal scenario where splitting is generated by extra term in the Lagrangian, like the one we have in our model.

\section{$\mu-\tau$ Symmetric Limit of the Model}
\label{sec:mutau}
In the most general case discussed above, the light neutrino mass matrix derived from the type I seesaw formula turns out to break $\mu-\tau$ symmetry resulting in non-zero $\theta_{13}$. The antisymmetric part of the triplet multiplications  $\frac{Y_a}{\Lambda} (\phi_{\nu} \bar{L})_{3_a} \tilde{H} N$ in the Dirac mass term is responsible for breaking the $\mu-\tau$ symmetry and in the limit of $ Y_a \rightarrow 0$, the $\mu-\tau$ symmetry in the light neutrino mass matrix can be recovered. In this limit, for the simple flavon vev alignment $\langle \phi_E \rangle =(v_E, 0, 0), \langle \phi_{\nu} \rangle =(v_{\nu}, v_{\nu}, v_{\nu})$, the charged lepton mass matrix is diagonal as before whereas the Dirac neutrino mass matrix takes a simpler form given by
\begin{equation}
M_D =  \left(\begin{array}{ccc}
 2a & -a & -a \\
       -a & 2a & -a \\ 
      -a & -a & 2a\end{array}\right)
\label{md3}
\end{equation}
where $a = \frac{v_H}{\Lambda} \frac{1}{3}Y_s v_{\nu}$. Using the right handed neutrino mass matrix given by \eqref{mr1}, the light neutrino mass matrix from type I seesaw formula can be written as
\begin{equation}
-M_{\nu} = M_D M^{-1}_R M^T_D= \frac{3a^2}{b} \left(\begin{array}{ccc}
2 & -1 & -1 \\
-1 & -1 & 2 \\
-1 & 2 & -1
\end{array}\right)
\label{mnu2}
\end{equation}
where $b=2 Y_N v_{\xi}$. This light neutrino mass matrix is clearly $\mu-\tau$ symmetric but it predicts two degenerate massive neutrinos and one massless neutrino, inconsistent with the observed mass squared differences.

We suitably modify the field content to arrive at a more realistic $\mu-\tau$ symmetric light neutrino mass matrix, as shown in table \ref{tab2}.
\begin{table}[htb]
\begin{tabular}{|c|c|c|c|c|c|c|c|c|c|c|c|}
\hline  & $ \bar{L} $ & $e_R$ & $\mu_R$ & $ \tau_R $ & $N$  & $H$ & $\phi_E$ & $\phi_{\nu}$ & $\xi$ & $\zeta$ & $\eta$\\
\hline
$SU(2)_L$ & 2 & 1 & 1 & 1 & 1 & 2 & 1 & 1 & 1 & 1 & 1 \\
$A_4$ & 3 & 1 & $1^{\prime}$ & $1^{\prime \prime}$ & 3 & 1 & 3 & 3 & 1 & $1^{\prime \prime}$ & 1 \\
$Z_3$ & $\omega$ & $\omega^2$ & $\omega^2$ & $\omega^2$ & $\omega$ & 1 & 1 & $\omega$ & $\omega$ & 1 & $\omega$ \\
$Z_2$ & 1 & 1 & 1 & 1 & -1 & 1 & 1 & -1 & 1 & -1 & -1\\ 
\hline     
\end{tabular}
\caption{Fields and their transformation properties under $ SU(2)_{L} $ gauge symmetry as well as the $ A_{4} $ symmetry in the $\mu-\tau$ symmetric limit.}
\label{tab2}
  \end{table} 
In the limit of vanishing antisymmetric part of the $A_4$ triplet products, the Yukawa Lagrangian for the Dirac neutrino mass terms can be written as
\begin{align}
\mathcal{L}_Y \supset \frac{Y_s}{\Lambda} (\phi_{\nu} \bar{L})_{3_s} \tilde{H} N + \frac{Y^{\prime}}{\Lambda} (\bar{L} N)_{1} \tilde{H} \eta + \text{h.c.}
\end{align}
In this case, the Dirac neutrino mass matrix can be written as
  \begin{equation}
     M_{D}= \left(\begin{array}{cccc}
      a+2b & -b & -b \\
       -b & 2b & a-b \\ 
      -b & a-b & 2b
      \end{array}\right).
      \label{mdmutau1}
      \end{equation}
where $b = \frac{v_H}{\Lambda} \frac{1}{3}Y_s v_{\nu}, a = \frac{v_H}{\Lambda} \frac{1}{3}Y^{\prime} v_{\eta}$, with $v_{\eta}$ being the vev of the flavon field $\eta$. Using the same leading order right handed neutrino mass matrix given by \eqref{mr1}, we can derive a $\mu-\tau$ symmetric light neutrino mass matrix using the type I seesaw formula. In fact, this gives rise to TBM type mixing, one of the widely studied neutrino mixing framework which was consistent with neutrino data prior to the discovery of non-zero $\theta_{13}$. Since the TBM can still be considered as a leading order approximation due to the smallness of $\theta_{13}$ compared to other mixing angles, such a scenario can be realistic provided a small deviation to it can be realised in order to generate non-zero $\theta_{13}$. This can be done simply by incorporating another flavon field $\psi$ that has the following transformation
$$ \psi (SU(2)_L: 1, \; A_4: 1^{\prime}, \; Z_3: \omega, \; Z_2: -1)$$
This allows one more contribution to Dirac neutrino mass term in the form of 
\begin{align}
\mathcal{L}_Y \supset \frac{Y^{\prime \prime}}{\Lambda} (\bar{L} N)_{1^{\prime \prime}} \tilde{H} \psi + \text{h.c.}
\end{align}
After the flavon field $\psi$ gets a vev $v_{\psi}$, this introduces a $\mu-\tau$ symmetry breaking correction to the Dirac mass term given by 
  \begin{equation}\
\delta M_{D} = \left(\begin{array}{cccc}
      0 & 0 & f \\
       0 & f & 0 \\ 
      f & 0 & 0
      \end{array}\right)
      \label{md:corr1}
      \end{equation}
where $f=\frac{v_H}{\Lambda} \frac{1}{3}Y^{\prime \prime} v_{\psi}$. Since this is a limiting case of the most general case based on an assumption of vanishing antisymmetric terms, we do not perform any numerical calculations for this scenario. The calculations will be similar to generic $A_4$ models where non-zero $\theta_{13}$ is generated by considering corrections to a leading order $\mu-\tau$ symmetric light neutrino mass matrix. For example, the work \cite{dbmkdsp} considered such a scenario.

\section{\label{sec:level4}Conclusion}
We have studied an extension of the standard model by discrete flavour symmetry $A_4 \times Z_3 \times Z_2$ that can simultaneously explain the correct neutrino oscillation data and the observed baryon asymmetry of the Universe. Considering a TeV scale type I seesaw we adopt the mechanism of resonant leptogenesis to generate a lepton asymmetry through out of equilibrium CP violating decay of right handed neutrinos which later gets converted into the required baryon asymmetry through electroweak sphalerons. The field content and its transformation under the flavour symmetry are chosen in such a way that the leading order right handed neutrino mass matrix has a trivial structure giving a degenerate spectrum. The tiny splitting between the right handed neutrino masses (required for resonant leptogenesis) arises through higher dimension mass terms, naturally suppressing the splitting. Due to the trivial structure of the right handed neutrino mass matrix, the leptonic mixing arises through the non-trivial structure of the Dirac neutrino mass matrix within a type I seesaw framework. This automatically leads to a $\mu-\tau$ symmetry breaking light neutrino mass matrix due to the existence of anti-symmetric terms arising from product of two triplet representations of $A_4$. Although such terms vanish for right handed neutrino mass matrix due to the Majorana nature, they do not vanish in general for Dirac neutrino mass matrix. Within a minimal setup, we then compare the $\mu-\tau$ symmetry breaking light neutrino mass matrix with the one constructed from light neutrino parameters and find the model parameters, while fixing the right handed neutrino mass at 5 TeV. Since there are only four independent complex parameters of the model that can be evaluated comparing four mass matrix elements, it gives rise to two constraints due to the existence of six independent complex elements of a light neutrino mass matrix which is complex symmetric if the light neutrinos are of Majorana type. These two constraints severely restrict the allowed parameter space to a narrow range, which we evaluate numerically by doing a random scan of ten million neutrino data points in the allowed $3\sigma$ range, for both normal and inverted hierarchical patterns of light neutrino masses. Among the unknown light neutrino parameters namely, the lightest neutrino mass, one Dirac and two Majorana CP phases, we get some interesting restrictions on some of these parameters from the requirement of satisfying the correct neutrino data within the model framework.

After finding the model and neutrino parameters consistent with the basic setup, we then feed the allowed parameters to the resonant leptogenesis formulas and calculate the baryon asymmetry of the Universe. We find that both the normal and inverted hierarchical scenarios can satisfy the Planck 2015 bound on baryon asymmetry $\eta_B = 6.04 \pm 0.08 \times 10^{-10}$ \cite{Planck15}. We however get more allowed points for the inverted hierarchical scenario compared to the normal one. Finally, we also briefly outline the $\mu-\tau$ symmetric limit of the model taking the approximation of vanishing anti-symmetric triplet product term and a possible way to generate non zero $\theta_{13}$ in that scenario. We however, do not perform any separate numerical calculation in this limiting scenario. We find it interesting that, just by trying to generate leptonic mixing through a non-trivial Dirac neutrino mass term automatically leads to broken $\mu-\tau$ symmetry, automatically generating non-zero $\theta_{13}$. This is in fact a more economical way to generate the correct neutrino oscillation data than taking the usual route of generating $\mu-\tau$ symmetric mass matrix at leading order followed by some next to leading order corrections responsible for generating $\theta_{13} \neq 0$ which was the usual procedure adopted after the discovery of non-zero $\theta_{13}$ in 2012. It is also interesting that the model can naturally generate the tiny mass splitting between right handed neutrinos and generate the required baryon asymmetry through the mechanism of resonant leptogenesis. Such TeV scale seesaw scenario can also have some other interesting implications in collider as well as rare decay experiments like lepton flavour violation, details of which can be found elsewhere. Also, such a TeV scale seesaw scenario can play a non-trivial role in restoring the electroweak vacuum stability as discussed recently by the authors of \cite{Bambhaniya:2016rbb}.
\appendix
\section{Details of $A_4$ Group}
\label{appen1}
$A_4$, the symmetry group of a tetrahedron, is a discrete non-abelian group of even permutations of four objects. It has four irreducible representations: three one-dimensional and one three dimensional which are denoted by $\bf{1}, \bf{1'}, \bf{1''}$ and $\bf{3}$ respectively, being consistent with the sum of square of the dimensions $\sum_i n_i^2=12$. We denote a generic permutation $(1,2,3,4) \rightarrow (n_1, n_2, n_3, n_4)$ simply by $(n_1 n_2 n_3 n_4)$. The group $A_4$ can be generated by two basic permutations $S$ and $T$ given by $S = (4321), T=(2314)$. This satisfies 
$$ S^2=T^3 =(ST)^3=1$$
which is called a presentation of the group. Their product rules of the irreducible representations are given as
$$ \bf{1} \otimes \bf{1} = \bf{1}$$
$$ \bf{1'}\otimes \bf{1'} = \bf{1''}$$
$$ \bf{1'} \otimes \bf{1''} = \bf{1} $$
$$ \bf{1''} \otimes \bf{1''} = \bf{1'}$$
$$ \bf{3} \otimes \bf{3} = \bf{1} \otimes \bf{1'} \otimes \bf{1''} \otimes \bf{3}_a \otimes \bf{3}_s $$
where $a$ and $s$ in the subscript corresponds to anti-symmetric and symmetric parts respectively. Denoting two triplets as $(a_1, b_1, c_1)$ and $(a_2, b_2, c_2)$ respectively, their direct product can be decomposed into the direct sum mentioned above. In the $S$ diagonal basis, the products are given as
$$ \bf{1} \backsim a_1a_2+b_1b_2+c_1c_2$$
$$ \bf{1'} \backsim a_1 a_2 + \omega^2 b_1 b_2 + \omega c_1 c_2$$
$$ \bf{1''} \backsim a_1 a_2 + \omega b_1 b_2 + \omega^2 c_1 c_2$$
$$\bf{3}_s \backsim (b_1c_2+c_1b_2, c_1a_2+a_1c_2, a_1b_2+b_1a_2)$$
$$ \bf{3}_a \backsim (b_1c_2-c_1b_2, c_1a_2-a_1c_2, a_1b_2-b_1a_2)$$
In the $T$ diagonal basis on the other hand, they can be written as
$$ \bf{1} \backsim a_1a_2+b_1c_2+c_1b_2$$
$$ \bf{1'} \backsim c_1c_2+a_1b_2+b_1a_2$$
$$ \bf{1''} \backsim b_1b_2+c_1a_2+a_1c_2$$
$$\bf{3}_s \backsim \frac{1}{3}(2a_1a_2-b_1c_2-c_1b_2, 2c_1c_2-a_1b_2-b_1a_2, 2b_1b_2-a_1c_2-c_1a_2)$$
$$ \bf{3}_a \backsim \frac{1}{2}(b_1c_2-c_1b_2, a_1b_2-b_1a_2, c_1a_2-a_1c_2)$$

\bibliographystyle{apsrev}

\begin{thebibliography}{99}
\expandafter\ifx\csname natexlab\endcsname\relax\def\natexlab#1{#1}\fi
\expandafter\ifx\csname bibnamefont\endcsname\relax
  \def\bibnamefont#1{#1}\fi
\expandafter\ifx\csname bibfnamefont\endcsname\relax
  \def\bibfnamefont#1{#1}\fi
\expandafter\ifx\csname citenamefont\endcsname\relax
  \def\citenamefont#1{#1}\fi
\expandafter\ifx\csname url\endcsname\relax
  \def\url#1{\texttt{#1}}\fi
\expandafter\ifx\csname urlprefix\endcsname\relax\def\urlprefix{URL }\fi
\providecommand{\bibinfo}[2]{#2}
\providecommand{\eprint}[2][]{\url{#2}}

\bibitem{PDG}
S.~Fukuda et al. (Super-Kamiokande),
{Phys. Rev. Lett.} {\bf 86}, 5656 (2001), hep-ex/0103033;
Q. R.~Ahmad et al. (SNO),
{Phys. Rev. Lett.} {\bf 89}, 011301 (2002), nucl-ex/0204008;
{Phys. Rev. Lett.} {\bf 89}, 011302 (2002), nucl-ex/0204009;
J. N.~Bahcall and C.~Pena-Garay,
{New J. Phys.} {\bf 6}, 63 (2004), hep-ph/0404061;
K. Nakamura et al., J.\ Phys.\ {\bf G37}, 075021 (2010).
\bibitem{kamland08}
S. Abe et al. [KamLAND Collaboration], Phys.Rev.Lett. {\bf 100}, 221803 (2008).
\bibitem{T2K} K. Abe et al. [T2K Collaboration],  Phys. Rev. Lett. {\bf 107}, 041801 (2011).
\bibitem{chooz} Y. Abe et al. [DOUBLE-CHOOZ Collaboration], Phys. Rev. Lett. {\bf 108}, 131801 (2012).
\bibitem{daya}  F. P. An et al. [DAYA-BAY Collaboration], Phys. Rev. Lett. {\bf 108}, 171803 (2012).
\bibitem{reno} J. K. Ahn et al. [RENO Collaboration], Phys. Rev. Lett. {\bf 108}, 191802 (2012).
\bibitem{minos}
P. Adamson et al. [MINOS Collaboration], Phys.Rev.Lett. {\bf 110}, 171801 (2013).
\bibitem{schwetz16} I. Esteban, M. C. Gonzalez-Garcia, M. Maltoni, I. Martinez-Soler and T. Schwetz, JHEP {\bf 1701}, 087 (2017).
\bibitem{valle17} P. F. de Salas, D. V. Forero, C. A. Ternes, M. Tortola and J. W. F. Valle, arXiv:1708.01186.
\bibitem{Planck15} P.~A.~R.~Ade et al., [Planck Collaboration], Astron. Astrophys. {\bf 594}, A13 (2016).
\bibitem{diracphase} K. Abe et al., [T2K Collaboration], Phys. Rev. {\bf D91}, 072010 (2015).
\bibitem{weinberg} S. Weinberg, Phys. Rev. Lett. {\bf 43}, 1566 (1979).
\bibitem{ti}
P.~Minkowski,
{Phys. Lett.} {\bf B67}, 421 (1977);
M.~Gell-Mann, P.~Ramond, and R.~Slansky (1980), print-80-0576 (CERN);
T.~Yanagida (1979), in Proceedings of the Workshop on the Baryon Number of the Universe and Unified Theories, Tsukuba, Japan, 13-14 Feb 1979;
R. N.~Mohapatra and G.~Senjanovic,
{Phys. Rev. Lett} {\bf 44}, 912 (1980);
J.~Schechter and J. W. F.~Valle,
{Phys. Rev.} {\bf D22}, 2227 (1980).
\bibitem{xing2015} Z. -z. Xing and Z. -z. Zhao, Rept. Prog. Phys. {\bf 79}, 076201 (2016).
\bibitem{Harrison} P.~F.~Harrison, D.~H.~Perkins and W.~G.~Scott, Phys. Lett. {\bf B530}, 167 (2002); P.~F.~Harrison and W.~G.~Scott, Phys. Lett. {\bf B535}, 163 (2002); Z.~z.~Xing, Phys. Lett. {\bf B533}, 85 (2002); P.~F.~Harrison and W.~G.~Scott, Phys. Lett. {\bf B547}, 219 (2002); P.~F.~Harrison and W.~G.~Scott, Phys. Lett. {\bf B557}, 76 (2003); P.~F.~Harrison and W.~G.~Scott, Phys. Lett. {\bf B594}, 324 (2004).
\bibitem{discreteRev} Y.~Shimizu, M.~Tanimoto and A.~Watanabe, Prog. Theor. Phys. {\bf 126}, 81 (2011); H. Ishimori, T. Kobayashi, H. Ohki, Y. Shimizu,
H. Okada and M. Tanimoto, Prog. Theor. Phys. Suppl.
183, 1 (2010); W. Grimus and P. O. Ludl, J. Phys. A 45, 233001 (2012); S. F. King and C. Luhn, Rept. Prog. Phys. 76, 056201; G. Altarelli and F. Feruglio, Rev. Mod. Phys. {\bf 82}, 2701 (2010).
\bibitem{A4TBM} E.~Ma and G.~Rajasekaran, Phys. Rev. {\bf D64}, 113012 (2001); E.~Ma, Mod. Phys. Lett. {\bf A17}, 627 (2002); K.~S.~Babu, E.~Ma and J.~W.~F.~Valle, Phys. Lett. {\bf B552}, 207 (2003); M.~Hirsch, J.~C.~Romao, S.~Skadhauge, J.~W.~F.~Valle and A.~Villanova del Moral, Phys. Rev. {\bf D69}, 093006 (2004); E.~Ma, Phys. Rev. {\bf D70}, 031901 (2004); E.~Ma, New J. Phys. {\bf 6}, 104 (2004); S.~L.~Chen, M.~Frigerio and E.~Ma, Nucl. Phys. {\bf B724}, 423 (2005); E.~Ma, Phys. Rev. {\bf D72}, 037301 (2005); A.~Zee, Phys. Lett. {\bf B630}, 58 (2005); E.~Ma, Mod. Phys. Lett. {\bf A20}, 2601 (2005); E.~Ma, Phys. Rev. {\bf D73}, 057304 (2006); S.~K.~Kang, Z.~z.~Xing and S.~Zhou, Phys. Rev. {\bf D73}, 013001 (2006).
\bibitem{A4TBM1} G.~Altarelli and F.~Feruglio, Nucl. Phys. {\bf B741}, 215 (2006).
\bibitem{nzt13} S.~F.~King and C.~Luhn, JHEP {\bf 1109}, 042 (2011); S.~Antusch, S.~F.~King, C.~Luhn and M.~Spinrath, Nucl. Phys. {\bf B856}, 328 (2012); S.~F.~King and C.~Luhn, JHEP {\bf 1203}, 036 (2012); S.~Gupta, A.~S.~Joshipura and K.~M.~Patel, Phys. Rev. {\bf D85}, 031903 (2012); S-F. Ge, D. A. Dicus and W. W. Repko, Phys. Rev. Lett. {\bf 108}, 041801 (2012); S-F. Ge, H-J. He and F-R. Yin, JCAP {\bf 1005}, 017 (2010); S-F. Ge, D. A. Dicus and W. W. Repko, Phys. Lett. {\bf B702}, 220 (2011); J. Liao, D. Marfatia and K. Whisnant, Phys. Rev. {\bf D87}, 013003 (2013); Z. -z. Xing, Phys. Lett. B 696, 232 (2011);
\bibitem{nzt13A4} B. Adhikary, B. Brahmachari, A. Ghosal, E. Ma and M. K. Parida, Phys. Lett. B 638, 345
(2006); E. Ma and
D. Wegman, Phys. Rev. Lett. 107, 061803 (2011) [arXiv:1106.4269 [hep-ph]]; G.~Altarelli, F.~Feruglio, L.~Merlo and E.~Stamou, JHEP {\bf 1208}, 021 (2012); B. Karmakar and A. Sil, Phys. Rev. {\bf D91}, 013004 (2015).
\bibitem{nzt13GA}  M-C.~Chen, J.~Huang, J-M.~O'Bryan, A.~M.~Wijangco and F.~Yu, JHEP {\bf 1302}, 021 (2012).
\bibitem{db-t2}D. Borah, Nucl. Phys. {\bf B876}, 575 (2013); D. Borah, S. Patra and P. Pritimita, Nucl. Phys. {\bf B881}, 444 (2014).
\bibitem{dbijmpa} D. Borah, Int. J. Mod. Phys. {\bf A29}, 1450108 (2014).
\bibitem{dbmkdsp} M. Borah. D. Borah, M. K. Das and S. Patra, Phys. Rev. {\bf D90}, 095020 (2014).
\bibitem{dbrk} R. Kalita and D. Borah, Int. J. Mod. Phys. {\bf A30}, 09, 1550045 (2015).
\bibitem{amdbmkd} A. Mukherjee, D. Borah and M. K. Das, Phys. Rev. {\bf D96}, 015014 (2017).
\bibitem{sakharov} A. Sakharov, Pisma Zh. Eksp. Teor. Fiz. {\bf 5}, 32 (1967).
\bibitem{davidsonPR} S. Davidson, E. Nardi and Y. Nir, Phys. Rept. {\bf 466}, 105 (2008).
\bibitem{fukuyana} M.~Fukugita and T.~Yanagida, Phys. Lett. {\bf B174}, 45 (1986).
\bibitem{sphaleron} V.~A.~Kuzmin, V.~A.~Rubakov and M.~E.~Shaposhnikov, Phys. Lett. {\bf B155}, 36 (1985).
\bibitem{davidsonibarra} S. Davidson and A. Ibarra, Phys. Lett. {\bf B535}, 25 (2002).

\bibitem{Pilaftsis:1997jf}
  A.~Pilaftsis,
  Phys.\ Rev.\ D {\bf 56} (1997) 5431
  [hep-ph/9707235].
\bibitem{Flanz:1996fb}
  M.~Flanz, E.~A.~Paschos, U.~Sarkar and J.~Weiss,
  Phys.\ Lett.\ B {\bf 389} (1996) 693
  [hep-ph/9607310].
\bibitem{Pilaftsis:2003gt}
  A.~Pilaftsis and T.~E.~J.~Underwood,
  Nucl.\ Phys.\ B {\bf 692} (2004) 303
  doi:10.1016/j.nuclphysb.2004.05.029
  [hep-ph/0309342].
\bibitem{Xing:2006ms}
  Z.~z.~Xing and S.~Zhou,
  Phys.\ Lett.\ B {\bf 653} (2007) 278
  [hep-ph/0607302].
\bibitem{dev2014} P. S. Bhupal Dev, P. Millington, A. Pilaftsis and D. Teresi, Nucl. Phys. {\bf B886}, 569 (2014).
\bibitem{Bambhaniya:2016rbb}
  G.~Bambhaniya, P.~S.~Bhupal Dev, S.~Goswami, S.~Khan and W.~Rodejohann,
  Phys.\ Rev.\ D {\bf 95} (2017) no.9,  095016
\bibitem{Kartavtsev:2015vto}
  A.~Kartavtsev, P.~Millington and H.~Vogel,
  JHEP {\bf 1606} (2016) 066.
  \bibitem{dev1410} P. S. Bhupal Dev, P. Millington, A. Pilaftsis and D. Teresi, Nucl. Phys. {\bf B891}, 128 (2015).
 \bibitem{Dev:2015wpa}
  P.~S.~B.~Dev, P.~Millington, A.~Pilaftsis and D.~Teresi,
  Nucl.\ Phys.\ B {\bf 897} (2015) 749
  \bibitem{Deppisch:2010fr}
  F.~F.~Deppisch and A.~Pilaftsis,
  Phys.\ Rev.\ D {\bf 83} (2011) 076007
\bibitem{radiative1} R. Gonzalez Felipe, F. R. Joaquim and B. M. Nobre, Phys. Rev. {\bf D70}, 085009 (2004); G. C. Branco, R. Gonzalez Felipe, F. R. Joaquim and B. M. Nobre, Phys. Lett. {\bf B633}, 336 (2006).
\bibitem{radiative2} A. Pilaftsis and T. E. J. Underwood, Phys. Rev. {\bf D72}, 113001 (2005).

   
\end{thebibliography}

\end{document}